\documentclass[amsmath,amssymb,amsbsy,reprint,prb,preprintnumbers,showpacs,superscriptaddress]{revtex4-2}
\usepackage{graphicx,color}
\usepackage{dcolumn}
\usepackage{bm}
\usepackage{braket}
\usepackage{mathtools}
\usepackage{ulem}
\usepackage[breaklinks,colorlinks=true,linkcolor=blue,urlcolor=blue,citecolor=blue]{hyperref}
\usepackage{times}
\usepackage{physics}
\usepackage{latexsym}
\usepackage{amsmath, amssymb}
\usepackage{mathtools}
\usepackage{multirow}

\newcommand{\be}{\begin{eqnarray}}
\newcommand{\ee}{\end{eqnarray}}

\renewcommand{\theequation}{\arabic{equation}}

\begin{document}
\title{Succession of Ising criticality and its threshold in critical quantum Ising model
subject to $X+ZZ$ decoherence}

\date{\today}
\author{Yoshihito Kuno} 
\affiliation{Graduate School of Engineering science, Akita University, Akita 010-8502, Japan}

\author{Takahiro Orito}
\affiliation{Department of Physics, College of Humanities and Sciences, Nihon University, Sakurajosui, Setagaya, Tokyo 156-8550, Japan}

\author{Ikuo Ichinose} 
\thanks{A professor emeritus}
\affiliation{Department of Applied Physics, Nagoya Institute of Technology, Nagoya, 466-8555, Japan}

\begin{abstract} 
We investigate a mixed state quantum criticality in the Ising model under $X+ZZ$ decoherence. 
In the doubled Hilbert space formalism, the decohered state resides on the self-dual critical line of the quantum Ashkin-Teller (qAT) model, as a result of the specific choice of the decoherence channel. 
On the other hand, since the mixed state under $X+ZZ$ decoherence satisfies the Kramers-Wannier self-duality in a weak sense, 
the Ising criticality of the pure state can be partially preserved in the mixed system. By making use of the combination of the doubled Hilbert space formalism and matrix product states, a numerical study is carried out. We numerically find that, up to moderate decoherence strength, the behavior of the mixed states on the critical line is consistent 
with Ising-like scaling within the accessible system sizes, with the effective central charge and critical exponents close to the Ising-CFT values $c=1/2$, $\eta=1/4$, and $\eta_X=2$. 
Our results of the central charge and critical exponents contradict the ones in the $c=1$ orbifold boson CFT describing the critical state of the qAT model. 
In addition, we also numerically observe the flow of the central charge up to strong decoherence. 
The strong decoherence washes out the remnant Ising criticality and induces strong-to-weak spontaneous symmetry breaking. 
\end{abstract}


\maketitle
\section{Introduction}
Effects of the environment \cite{gardiner2000,Zurek2003decoherence} on quantum systems are inevitable. 
Many-body quantum systems, including ones of great practical use, such as quantum computers and quantum memories, 
are significantly affected by undesired noise and decoherence. 
Then, the target many-body states are destroyed by noise and/or decoherence when they exceed a threshold~\cite{preskill2018,dennis2002,wang2003,ohno2004}. 
However, even if such noise and decoherence change the target state to a mixed state, losing the original coherence, some quantum orders of the pure state are to be sustained in the mixed state. 
For example, the topological order of the toric code \cite{ebadi2021,bluvstein2024}
is expected to perfectly or partially survive even when the state gets mixed by noise or decoherence,
as quantum or sometimes classical memories \cite{Fan_2024,wang2024,sohal2024,KOI2024_IMTO}. 
These examples indicate not only the robustness of pure states but also the richness of novel mixed states emerging from interactions with the environment. 
That is, the effect of noise and decoherence can lead to rich non-trivial quantum mixed states never being created in isolated quantum systems. 
Noise and decoherence applied to pure quantum states can be an essential ingredient to produce exotic mixed quantum states, which can play an important role in quantum devices. 

Along this current research trend, how critical quantum many-body states are affected by various types of decoherence is an on-going issue. We have not completely understood it nor arrived to established consensus. 
In particular, how decoherence or noise changes physical and information properties of critical states described by conformal field theory (CFT) \cite{di1996conformal,Henkel1999conformal}
is an important issue. 
The fate of the universal property of the well-studied CFTs under decoherence should be investigated in detail.

As a first step toward elucidating this global issue, we study the simplest pure critical state subject to a specific type of decoherence, that is, 
the critical ground state of the transverse field Ising model (TFIM) described by Ising CFT \cite{di1996conformal,Henkel1999conformal}.
Here, we consider $X+ZZ$ type decoherence. 
This decoherence has particular properties since the corresponding operation can be regarded as the part of the inter-leg interactions of the quantum Ashkin-Teller (qAT) model 
in the doubled Hilbert space formalism. 
Then, the Kraus operators of the decoherence channel possess a Kramers–Wannier (KW) symmetry in a weak sense. 

Previous studies \cite{Zou2023,KOI2025_6} have investigated the critical ground state of the transverse field Ising model (TFIM) subject to 
local Pauli decoherence. 
It was shown that the CFT subsystem scaling survives. 
The studies ~\cite{Zou2023,KOI2025_ZZ_deco} also showed that the universality class of the criticality varies depending on the individual decoherence channel. 
At present, the mechanism of this nontrivial behavior of critical states under decoherence is poorly understood from universal viewpoint. 
Building on these prior studies, this work focuses on the TFIM subject to the $X+ZZ$ decoherence, 
as the resulting mixed state in the doubled Hilbert space formalism is expected to be closely related to the ground states in the qAT model~\cite{Kohmoto1981,Solyom,Yamanaka1994,O'Brien2015,Bridgeman2015}. 

This work numerically investigates the criticality of the mixed states on the critical line of the phase diagram.
In particular, we numerically investigate how the CFT properties of the critical state of the TFIM change or survive under the aforementioned decoherence, 
with focusing on some physical quantities such as the central charge and critical exponents of correlation functions.
Practically, we utilize numerical filtering operation methods applying to the matrix product state (MPS). 
By this numerical scheme, we calculate the R\'{e}nyi-2 subsystem entanglement entropy (EE) of mixed states and observe its subsystem dependence in detail. 

In general, the R\'{e}nyi-2 subsystem EE of the mixed state does not agree with that of the pure state~\cite{Zou2023} 
since the scaling of the R\'{e}nyi-2 subsystem EE can exhibit a different form in the leading term of the system size for pure and mixed states. But it has the same CFT scaling form for the subleading term in both states. 
Thus, we focus on the subscaling term in the decohered system. 
According to Ref.~\cite{Zou2023}, the scaling form of the subleading term can be elucidated by introducing the R\'{e}nyi-2 mutual information (R2MI) 
composed of the R\'{e}nyi-2 subsystem EE.
Its scaling form obeys the CFT scaling of R2MI given as,
$$
\frac{c_{2}}{4} \ln\left[ \frac{L}{\pi} \sin\left( \frac{\ell \pi}{L} \right) \right] + \gamma,
$$
where $L$ is the system size (we employ periodic boundary conditions), 
$\ell$ is subsystem size, $c_2$ is R\'{e}nyi-2 central charge, and $\gamma$ is non-universal constant.
In particular, there exists a conjecture \cite{CC2004,Stephan2010,Alcaraz2013,Alcaraz2014,Stephan2014}; $c_2=2c_{\rm eff}$, where $c_{\rm eff}$ is an effective central charge, 
and for Ising CFT states, $c_{\rm eff}=1/2$, that is, the central charge of the Ising CFT \cite{di1996conformal}. 
This conjecture has been numerically tested in the projective measurement limit of the critical state of the TFIM \cite{Stephan2010,Alcaraz2013,Alcaraz2014,Stephan2014}, 
to find the robustness of the CFT scaling and the central charge. 
However, this robustness of the CFT properties under decoherence has been observed for particular cases and 
there is no rigorous proof even for some specific cases. Thus, further numerical investigation is desired.

Through the known phase diagram of the qAT model, 
the phase diagram of the target decohered state can be inferred, and the previous work \cite{OKI2025_2} confirmed the presence of various mixed state phases;
counterparts of those in the qAT model. 
But the criticality has not been elucidated in detail. 
In this work, we investigate how the criticality of the TFIM evolves under $X+ZZ$ decoherence by observing the R\'{e}nyi-2 subsystem EE, etc. 
In our decoherence setup, the weak KW self-duality is kept. This gives an expectation of the survival of the Ising CFT.
Through extensive numerical calculations, we find that the critical
properties of the mixed states on the phase-boundary line are
consistent with Ising-like critical scaling within the accessible
system sizes, up to moderate strength of $X+ZZ$ decoherence. 
Further, we observe the threshold of the CFT property and show the emergence of the Z2 strong-to-weak spontaneous symmetry breaking (SWSSB) phase.

The rest of this paper is organized as follows. 
In Sec.~II, we explain the setting of the system in this work and consider the effects of the $X+ZZ$ decoherence and qualitative relationship to the qAT model.
In Sec.~III, we introduce quantum entanglement measures and some physical quantities to evaluate the criticality of the mixed state. 
Then, efficient numerical calculation methods are reviewed, where we introduce the doubled Hilbert space formalism.
In Sec.~IV, we display results of numerical calculations by using the MPS and the filtering to the MPS for various decoherence parameters. 
Section V is devoted to the conclusion and discussion.

Here, we give additional comments on the usage of the numerics of the doubled Hilbert space formalism: 
(I) We employ this formalism for numerical calculation to evaluate the criticality of the mixed states.
However, to date, a study demonstrating that exponents obtained by the doubled Hilbert space formalism for decohered mixed states
agree with known CFTs such as Ising, orbifold ones is lacking. 
Thus, this work provides a concrete numerical demonstration for the estimation of the criticality by the doubled Hilbert space formalism. 
(II) In the doubled-Hilbert-space representation, mixed-state system of $L$-site chain is mapped onto a two-leg ladder with $2L$ spin-$1/2$ degrees of freedom.  
In the present study, we access system sizes up to $L=40$, corresponding to an $80$-site ladder (with $3L=120$ interaction bonds under periodic
boundary conditions). Our aim is not to provide a high-precision determination of the asymptotic critical exponents, but rather to examine whether 
several independent observables exhibit mutually consistent Ising-like critical scaling within the accessible system sizes. 

\section{Target system under decoherence}

\subsection{Model and decoherence}
This work studies the 1D TFIM, the Hamiltonian of which is given by,
\begin{eqnarray}
H_{0}=-\sum^{L-1}_{j=0}[JZ_jZ_{j+1}+hX_j],\nonumber
\end{eqnarray}
where $J$ and $h$ are model parameters and periodic boundary conditions (PBCs) 
are imposed. 
The system possesses $Z_2$ symmetry, the generator of which is a global spin flip $\prod^{L-1}_{j=0}X_{j}$. 
At $J=h$, a critical ground state appears and is described by the Ising CFT\cite{di1996conformal,Henkel1999conformal}. 
For $J/h>1$, the ground state is $Z_2$ SSB ferromagnetic (FM) and for $J/h<1$, a paramagnetic (PM) state emerges. 
Hereafter, we denote the ground state of $H_0$ by $|\psi_0\rangle$, and its (pure-state) density matrix by $\rho_0=|\psi_0\rangle\langle \psi_0|$,
and $\mathcal{H}$ is the Hilbert space of the spin-$1/2$ $L$-site system.
 
In this work, we consider two types of tunable decoherence channels given as \cite{Nielsen2011}
\begin{eqnarray}
\mathcal{E}_{ZZ}[\rho]&=&\prod^{L-1}_{j=0}\mathcal{E}_{ZZ,j}[\rho],\nonumber\\
\mathcal{E}_{ZZ,j}[\rho]&=&\biggr[(1-p_{zz})\rho+p_{zz}Z_{j}Z_{j+1}\rho Z_{j+1}Z_{j}\biggl],\nonumber\\ 
\mathcal{E}_{X}[\rho]&=&\prod^{L-1}_{j=0}\mathcal{E}_{X,j}[\rho],\nonumber\\
\mathcal{E}_{X,j}[\rho]&=&\biggr[(1-p_{x})\rho+p_{x}X_{j}\rho X_{j}\biggl], \nonumber
\end{eqnarray} 
where $p_{zz(x)}$ is the strength of the decoherence and $0\leq p_{zz(x)}\leq 1/2$. 
The two types of decoherence are uniformly applied to the whole system as \cite{Note_deco} 
$$
\rho_D\equiv \mathcal{E}_{ZZ}\circ \mathcal{E}_{X}[\rho_0].
$$
In the previous work of the present authors \cite{OKI2025_2}, the phase diagram of the mixed state $\rho_D$ is clarified as in Fig.~\ref{Fig1}, 
in which there exist three phases, mixed FM, mixed PM, and SWSSB phase \cite{OKI2025_2}. There, phase boundaries were obtained by studying spin correlation functions and entanglement of the doubled system in the doubled Hilbert space formalism.
More details are explained in the following subsection after introducing the doubled Hilbert space formalism.

This work aims to clarify the behavior of the state on the phase boundary between the mixed FM and mixed PM. 
The condition $J=h$ fixes the self-dual point of the underlying noiseless TFIM, while $p_x$ and $p_{zz}$ are chosen to satisfy the effective self-dual critical-line condition of the corresponding qAT model. 
Under these conditions, the combined decoherence channel $\mathcal{E}_{ZZ}\circ\mathcal{E}_{X}$ can be written as
\begin{eqnarray}
\mathcal{E}_{D}[\rho]&=&\prod^{L-1}_{j=0}\mathcal{E}^{j}_D[\rho],\\
\mathcal{E}^{j}_D[\rho]&=&\sum_{\ell=0,1,2,3}K^{\ell}_{j}\rho K^{\ell,\dagger}_{j},
\end{eqnarray} 
where $K^{\ell=0,1,2,3}_j$ is a Kraus operator given by $K^{0}_j=(1-p_D)I$, $K^{1}_j=\sqrt{p_D(1-p_D)}X_j$, $K^{2}_j=\sqrt{p_D(1-p_D)}Z_jZ_{j+1}$ and $K^{3}_j=p_DX_j Z_jZ_{j+1}$ 
with a condition $\sum_{\ell}K^{\ell}_jK^{\ell,\dagger}_{j}=I$, $p_{D}\equiv p_{zz(x)}$ and 
the strength of the decoherence is tuned by $p_{D}$, taking $0\leq p_{D}\leq 1/2$. 
For $p_{D}= 1/2$, these channels correspond to projective measurements of $Z_jZ_{j+1}$ and $X_j$ without monitoring and are called maximal decoherence. 
It is easily verified that the decoherence $\mathcal{E}_{D}$ is weakly symmetric under the KW transformation, $Z_jZ_{j+1}\longrightarrow X_j$ and $X_j\longrightarrow Z_jZ_{j+1}$ 
if we ignore rigorous consideration of boundary constraints \cite{LinhaoLi2025}. 
In fact, each $K^{\alpha}_{j}$ is transformed to its pair Kraus operator under the KW transformation, showing weak symmetry of the decoherence channel \cite{groot2022,Guo-and-Ashida2024}.

\subsection{Doubled Hilbert space formalism and qualitative connection to quantum Ashkin-Teller model}
The qualitative phase structure of the decohered state $\rho_D$ can be understood from the doubled Hilbert space formalism of the present system. 
We briefly review this formalism and show that the mixed phase diagram can be over-viewed from the qAT model \cite{Ashkin1943,Solyom}. 

To explain the doubled Hilbert space formalism, we consider a state in a Hilbert space, $\rho\in \mathcal{H}$.
In the doubled Hilbert space formalism, the target Hilbert space is doubled as $\mathcal{H}_{u}\otimes \mathcal{H}_{\ell}$, 
where the subscripts $u$ and $\ell$ denote the upper and lower Hilbert spaces corresponding to ket and bra states of mixed state density matrix, respectively. 
A density matrix $\rho$ can be vectorized $\rho \longrightarrow |\rho\rangle\rangle$ \cite{Choi1975,JAMIOLKOWSKI1972} 
as $|\rho\rangle\rangle\equiv \frac{1}{\sqrt{\dim[\rho]}}\sum_{k}|k\rangle\otimes \rho|k\rangle$, 
where $\{|k\rangle \}$ is an orthonormal set of bases in the Hilbert space $\mathcal{H}$. 
Then, the state $|\rho\rangle\rangle$ is in the doubled Hilbert space $\mathcal{H}_u\otimes \mathcal{H}_{\ell}$.

In this formalism, decoherence channels can be mapped into an operator acting on the vector $|\rho\rangle\rangle$~\cite{lee2023,Lee2024}. 
Concretely, the Kraus operator sum of a decoherence represented by $\mathcal{E}[\rho]=\sum^{M-1}_{\alpha=0}K_{\alpha}\rho K^\dagger_{\alpha}$, 
where $K_\alpha$'s are Kraus operators, is turned into an operator in the space 
$\mathcal{H}_u\otimes \mathcal{H}_{\ell}$, given by $\hat{\mathcal{E}}=\sum^{M-1}_{\alpha=0}K^*_{\alpha,u}\otimes K_{\alpha,\ell}$.
The action of the decoherence is performed as $\hat{\mathcal{E}}|\rho\rangle\rangle$, which is a decohered state in the doubled Hilbert space. 

We consider the target system from the view of the doubled Hilbert space formalism. 
We set the input density matrix $\rho_0$ to be the ground state of the TFIM, denoted by $|\rho_0\rangle\rangle \equiv |\psi^{*}_0\rangle|\psi_0\rangle$, where the superscript $*$ denotes complex conjugation in the computational ($Z$) basis. 
The state $|\rho_0\rangle\rangle$ can be regarded as two copies of the ground state of the 1D TFIM $|\psi_0\rangle$. 
The doubled system then is a two-decoupled TFIM on a two-leg spin-$1/2$ ladder based on the doubled Hilbert space $\mathcal{H}_{u}\otimes \mathcal{H}_{\ell}$. 

Next, in this formalism, the decoherence channel $\mathcal{E}_{ZZ}\circ \mathcal{E}_{X}$ acting on $|\rho_0\rangle\rangle$ can be treated as a filtering operator $\hat{\mathcal{E}}_{ZZ}\hat{\mathcal{E}}_{X}$. 
The decohered state is given by
\begin{eqnarray}
|\rho_D\rangle\rangle&\equiv& \hat{\mathcal{E}}_{ZZ}\hat{\mathcal{E}}_{X}|\rho_0\rangle\rangle\propto \prod^{L-1}_{j=0}\biggr[e^{\tau_{zz} \hat{h}^{zz}_{j,j+1}}e^{\tau_{x} \hat{h}^x_{j}}\biggl]|\rho_0\rangle\rangle,
\label{rhoD}
\end{eqnarray}
where $\hat{h}^{zz}_{j,j+1}=Z_{j,u}Z_{j+1,u}\otimes Z_{j,\ell}Z_{j+1,\ell}$, $\hat{h}^x_j=X_{j,u}\otimes X_{j,\ell}$, $\tau_{zz(x)}=\tanh^{-1}[{p_{zz(x)}/(1-p_{zz(x)})}]$. 
In general, the state $|\rho_D\rangle\rangle$ is not a normalized state~\cite{Coef_nomal}. 
Equation (\ref{rhoD}) can be viewed that the state $|\rho_0\rangle\rangle$ is locally-filtered by the two different kinds of local perturbative operators
$e^{\tau_{zz} \hat{h}^{zz}_{j,j+1}}$ and $e^{\tau_{x} \hat{h}^x_{j}}$, which is analogue to the imaginary time evolution emerging the state $|\rho_D\rangle\rangle$~\cite{Ardonne2004,CASTELNOVO2005,Castelnovo2008,Haegeman2015}. 

\begin{figure}[t]
\begin{center} 
\vspace{0.5cm}
\includegraphics[width=6cm]{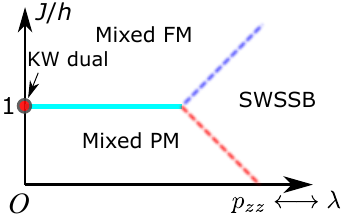}  
\end{center} 
\caption{Schematic phase diagram of mixed states. 
The existence of each mixed phase has been clarified in the previous work \cite{OKI2025_2}. 
In our numerical calculation, the strength of decoherence, $p_{zz}$ and $p_x$ is related to $\lambda$ in the qAT model as $\lambda \longleftrightarrow \tanh^{-1}[{p_{zz(x)}/(1-p_{zz(x)})}]$ 
with the constraint $p_x=1/2-(1/2)(1-2p_{zz})^{1/J}$. 
The global mixed phase diagram of the doubled system is closely related to that of the qAT model~\cite{Kohmoto1981}. 
The right blue line is a critical phase boundary line expected in \cite{OKI2025_2}. The red point is the KW duality point of the TFIM.}
\label{Fig1}
\end{figure}
Based on the previous studies \cite{Castelnovo2008,Haegeman2015,Zhu2019,Chen2024_v2}, this perturbative picture of the decoherence operation gives an effective Hamiltonian picture in the doubled system regarded as the spin-$1/2$ ladder system. The decoherence operation adds effective interactions to the doubled TFIM Hamiltonian of the ladder system. Then, we expect that the decohered state is qualitatively close to the ground state of the TFIM Hamiltonian with the effective interactions \cite{Filtering}, where the strengths of the effective interaction are proportional to $\tau_{zz}$ and $\tau_{x}$ tuned by $p_{zz}$ and $p_x$. 
The effective Hamiltonian of which is given on the ladder as the qAT model \cite{Kohmoto1981},
\begin{eqnarray}
H_{qAT}&=&-J\sum^{L-1}_{j=0}[Z_{j,u}Z_{j+1,u}+Z_{j,\ell}Z_{j+1,\ell}\nonumber\\
&&+\lambda_{zz} Z_{j,u}Z_{j,\ell}Z_{j+1,u}Z_{j+1,\ell}]\nonumber\\
&&-h\sum^{L-1}_{j=0}[X_{j,u}+X_{j,\ell}+\lambda_x X_{j,u}X_{j,\ell}].\nonumber
\end{eqnarray}
The global ground state phase diagram of $H_{qAT}$ has been already investigated in detail ~\cite{Kohmoto1981,O'Brien2015,Bridgeman2015}.  
There are three representative ground-state phases for the region $\lambda_{zz}=\lambda_{x}\equiv \lambda>0$ (since $\tau_{zz(x)}>0$) \cite{Kohmoto1981,Yamanaka1994}: 
Phase I. Double chain spontaneous $Z_2$ symmetry broken phase, Phase II Double paramagnetic phase. Phase III. Diagonal $Z_2$ symmetric phase (also called ``partially-ordered phase"). We also note that similar mappings between the transverse-field Ising chain and the qAT model in the doubled Hilbert space have also been discussed previously \cite{PhysRevB.99.224432}.

The previous study of the authors \cite{OKI2025_2} clarified that the decohered state $|\rho_D\rangle\rangle$ exhibits three mixed phases related to 
the above three phases in the qAT model: Phases I, II and III correspond to mixed FM phase, mixed PM and SWSSB phases, respectively. 
A schematic phase diagram is shown in Fig.~\ref{Fig1}. 
The phase boundary between phases I and II is obtained by the standard spin-spin correlation function, whereas phase III can be recognized by
the spin-spin correlation function in the doubled Hilbert space (see later discussion).
The EE also exhibits anomalous behavior, indicating transitions between the three phases. 
However, the detailed knowledge of the criticality of the mixed system is still lacking, and therefore, we shall study it in detail in this work.

For the qAT model, the criticality of the phase boundary $J/h=1$ was numerically investigated in detail \cite{Bridgeman2015} (pure state framework). 
The $\lambda$-term in $H_{qAT}$ is a marginal perturbation, inducing a continuous change of the criticality, the image of which is shown by the right-blue line in Fig.~\ref{Fig1}. The criticality holds on the line, $-1/\sqrt{2}\leq \lambda \leq 1$ and is described by the orbifold boson CFT \cite{di1996conformal}.

Here, a question arises whether the orbifold boson CFT of the qAT model is an effective field theory for the critical state under the $X+ZZ$ decoherence. 
In other words, we would like to elucidate how the criticality of the system changes under decoherence, as we explained in Sec.I.
In what follows, we study this issue by investigating the subsystem R\'{e}nyi-2 entanglement entropy and correlation functions on the critical line.

\section{Physical observables}
To elucidate the criticality of the mixed state, we introduce some physical quantities. 
The first one is the subsystem R2MI, defined as 
\begin{eqnarray}
MI^{(2)}(L_A)\equiv S^{(2)}_{A}+S^{(2)}_{B}-S^{(2)}_{A\cup B},\label{R2MI}
\end{eqnarray}
where the system is divided into two halves, denoted by the subsystem $A$ and $B$, and $A\cup B$ represents the entire system. 
$L_A$ is the subsystem size of the subsystem $A$. Here, $S^{(2)}_{X}$ is defined as $S^{(2)}_{X}=-\log \Tr[\rho^2_{D,X}]$ where $\rho_{D,X}\equiv \Tr_{\bar{X}}[\rho_{D}]$. 
The practical calculation scheme in the doubled Hilbert space formalism has been developed \cite{KOI_2025_v4} as briefly explained in Appendix A.

The second and third observables are the (reduced) susceptibilities of the R\'{e}nyi-2 $ZZ$ and conventional $ZZ$ correlators, 
\begin{eqnarray}
\chi^{\rm II}&=&{2 \over L}\sum^{L/2}_{r=1}C^{\rm II}_{ZZ}(0,r),\\
\chi^{\rm I}&=&\frac{2}{L}\sum^{L/2}_{r=1}C^{\rm I}_{Z}(0,r).
\label{chaiIIs}
\end{eqnarray}
Here, 
\begin{eqnarray}
C^{\rm II}_{ZZ}(i,j)&\equiv& \frac{\langle\langle \rho_D|Z_{i,u}Z_{j,u}Z_{i,\ell}Z_{j,\ell}|\rho_D\rangle\rangle}{\langle\langle \rho_D|\rho_D\rangle\rangle},\nonumber\\
C^{\rm I}_{Z}(i,j)&\equiv&\frac{\langle\langle {\bf 1}|Z_{i,u}Z_{j,u}|\rho_D\rangle\rangle}{\langle\langle {\bf 1}|\rho_D\rangle\rangle}.
\label{CIICI}
\end{eqnarray}
where $|{\bf 1}\rangle\rangle\equiv \displaystyle{\frac{1}{2^{3L/2}}\prod^{L-1}_{j=0}|t\rangle_j}$ with $|t\rangle_j=|\uparrow_u\uparrow_{\ell}\rangle_j+|\downarrow_u\downarrow_{\ell}\rangle_j$. 
In the original physical 1D system, $C^{\rm II}_{ZZ}(i,j)$ corresponds to the R\'{e}nyi-2 correlator;
$\displaystyle{C^{\rm II}_{ZZ}(i,j)\equiv \frac{\Tr[Z_iZ_j\rho_D Z_jZ_i \rho_D]}{\Tr[(\rho_D)^2]}}$, identifying the spontaneous symmetry breaking (SSB) of strong symmetry but {\it not} that of the weak one~\cite{lee2023,lessa2024,sala2024}.
$C^{\rm II}_{ZZ}(i,j)\neq 0$ for $|i-j| \to \infty$.
indicates the emergence of an SSB state. 
$C^{\rm I}_{Z}(i,j)$ in the original physical Hilbert space is given by ${\rm Tr}[\rho_D Z_{i}Z_{j}]$, which is the canonical two point $ZZ$ correlation, 
diagnosing the weak SSB.
More explanation about $C^{\rm I}_{Z}(i,j)$ is given in Appendix B.
The quantities $\chi^{\rm II}$ and  $\chi^{\rm I}$ are used as order parameters of these SSBs.
The combination of $\chi^{\rm II}$ and $\chi^{\rm I}$ can detect the SWSSB, 
which is recently proposed in Refs.~\cite{lee2023,lessa2024,sala2024,OKI2025_2} for strong symmetric systems.
In the doubled Hilbert space formalism, a state with $\chi^{\rm II}\sim \mathcal{O}(1)$ and $\chi^{\rm I}\sim 0$ exhibits SSB of the off-diagonal (i.e., strong) symmetry 
and also the restoration of the diagonal (i.e., weak) symmetry~\cite{lee2023}.

Here, we comment on weak KW self-duality in order.
A weak KW self-dual density matrix $\rho$ satisfies the following equation,
$$
(KW)\rho(KW) =\rho,
$$
where $(KW)$ schematically stands for the KW transformation. It is easily verified that the relation between the R\'{e}nyi-2 correlations diagnoses if the state has the weak KW symmetry;
\begin{eqnarray}
\mbox{Tr}[Z_iZ_j \rho Z_jZ_i \rho]= \mbox{Tr}\biggl[(\prod_{i\leq\ell<j} X_\ell)\rho (\prod_{i\leq\ell<j} X_\ell) \rho\biggr],
\end{eqnarray}
where the quantity on the RHS is nothing but the disorder parameter of weak $Z_2$ symmetry.
If the above equality holds, the state keeps the weak KW symmetry. 
We numerically verified the above equality for the states on the line $J=h$ and for a finite $p_{zz}$ regime. (See Appendix C). 
This fact implies that the critical mixed states with the weak KW duality can possess the properties of the Ising CFT since, as a broad consensus, the pure critical state
with the KW duality in the TFIM possesses the properties of the Ising CFT.

\section{Numerical investigation}
In this section, we numerically investigate states on the phase boundary line between the mixed PM and FM phases. 
The parameter sweep is performed by varying $p_{zz}$ with $p_x=1/2-(1/2)(1-2p_{zz})^{1/J}$. 
In most cases, we focus on the case $J=h=1$.

We generate the decohered state $|\rho_D\rangle\rangle$ by using the MPS formalism, which can analyze large ladder systems and 
clarify subsystem entanglement of the decohered state $|\rho_D\rangle\rangle$. 
For the numerical treatment of the MPS, we employ the TeNPy library \cite{TeNPy,Hauschild2024}.

First, we prepare the initial state $|\rho_0\rangle\rangle$, which is the ground state of the decoupled 1D TFIM on the ladder by the density matrix renormalization group algorithm. Then, 
the decoherence operation [the filtering operation in Eq.~(\ref{rhoD})] $\hat{\mathcal{E}}_{ZZ}\hat{\mathcal{E}}_{X}|\rho_0\rangle\rangle$ is efficiently carried out by the TeNPy package.  

In all the following numerical simulations, we treat the MPS with maximum bond dimension $D=300$, where singular values less than $\mathcal{O}(10^{-6})$ are truncated.   
In the density matrix renormalization group algorithm, the energy convergence of the iterative sweeping is $\Delta E < \mathcal{O}(10^{-5})$ to obtain the initial MPS ground state.
We then obtain the state $|\rho_D\rangle\rangle=\hat{\mathcal{E}}_{ZZ}\hat{\mathcal{E}}_{X}|\rho_0\rangle\rangle$ for various values of $p_{zz}$. 
In particular, the choice of the maximum bond dimension $D=300$ is sufficiently large
to get the convergent results in the following numerics. Due to the finite accessible system sizes, the numerical analyses presented below are not intended to yield a highly precise determination of critical exponents in the thermodynamic limit. 
Instead, we examine whether the observed scaling behaviors and extracted exponents are consistent with those of the Ising CFT within the available system sizes
by carefully monitoring finite-size effects.

\subsection{Subsystem R\'{e}nyi-2 mutual information}
We observe the subsystem R2MI.
As the previous works \cite{AFFLECK199735,Zou2023} showed, the subsystem R\'{e}nyi-2 EE exhibits the following scaling form if the system is described by a CFT 
\begin{eqnarray}
S^{(2)}_A(L_A)=\alpha_{0}L_A+\frac{c_{\rm eff}}{4}\log \biggr[\frac{L}{\pi}\sin\biggl(\frac{\pi L_A}{L}\biggr)\biggr]+\alpha_{1}.
\label{S2_con}
\end{eqnarray}
This form is valid for $0<L_A<L$. 
The subsystem EE has an extensive term proportional to $L_A$, but the sub-leading sine function term is universal, and $c_{\rm eff}$ is
called an effective central charge whereas the coefficient $\alpha_0$ is nonuniversal. 
As the above scaling form has not been derived analytically in our system, we shall numerically investigate the presence of the sub-leading term
to confirm the scaling law (\ref{S2_con}) and extract the effective central charge. 
\begin{figure}[t]
\begin{center} 
\vspace{0.5cm}
\includegraphics[width=9cm]{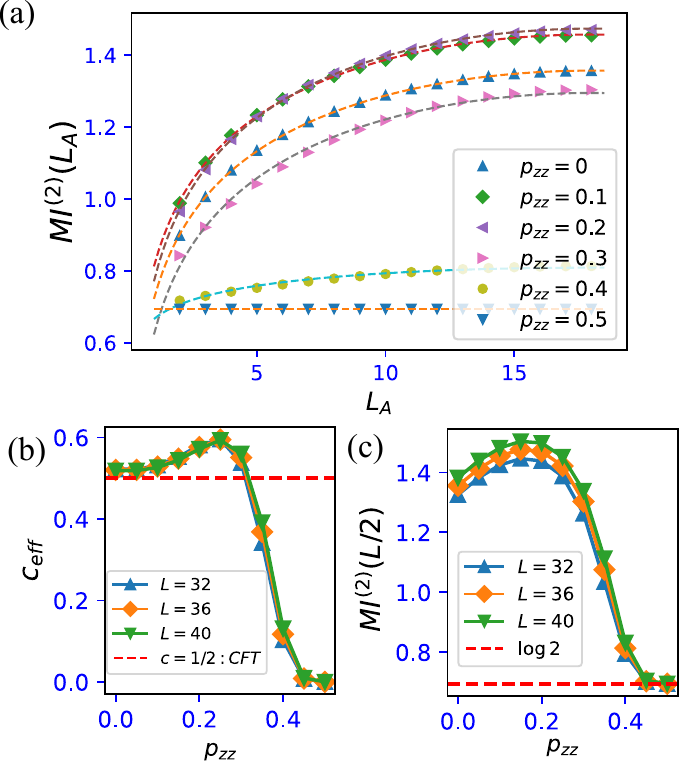}  
\end{center} 
\caption{(a) Subsystem size dependence of R2MI for various $p_{zz}$'s.
We set $L=40$. The dashed lines are the extracted fitting lines of the form of Eq.(\ref{fit_MI}).  
(c) $p_{zz}$-dependence effective extracted central charge for different system sizes. (b) $p_{zz}$-dependence of R2MI for $L_A=L/2$. 
For all data, we set $J=h=1$.}
\label{Fig2}
\end{figure}
In the following, we numerically examine if the R2MI satisfies the scaling law derived from Eq.~(\ref{S2_con}) \cite{Zou2023},  
\begin{eqnarray}
MI^{(2)}(L_A)=\frac{c_{\rm eff}}{2}\log \biggr[\frac{L}{\pi}\sin\biggl(\frac{\pi L_A}{L}\biggr)\biggr]+\beta_{1},
\label{fit_MI}
\end{eqnarray}
where $\beta_1$ is a non-universal constant \cite{footnote1}. 
The reason why we use the R2MI instead of the EE is that $c_{\rm eff}$ can be precisely estimated by the numerical optimization procedure,
eliminating the non-universal extensive term with the coefficient $\alpha_0$ in $S^{(2)}_A(L_A)$ [Eq.~(\ref{S2_con})].

Figure 2 (a) shows numerical results of the $L_A$ dependence of the R2MI for various values of $p_{zz}$. 
We find that the clear CFT scaling (sine function form) appears up to $p_{zz}=0.3$ and for $p_{zz}\gtrsim 0.4$, from the fitting with Eq.~(\ref{fit_MI}), $c_{\rm eff}$ takes smaller values than $1/2$ though the scaling form is not substantially disrupted. 
These results indicate that Ising-like critical scaling persists
up to $p_{zz}=0.3$ within the accessible system sizes.

Next, we numerically estimate the effective central charge $c_{\rm eff}$ from the data shown in Fig.~\ref{Fig2}(b). 
We confirmed that at $p_{zz}=0$ (without the decoherence) $c_{\rm eff}\sim 1/2$ consistent with the central charge of the Ising CFT based on the expectation for 
the R\'{e}nyi-2 EE \cite{CC2004,Stephan2010,Alcaraz2013,Alcaraz2014,Stephan2014}. 
Then we find that for moderate values of $p_{zz}$, the data are significantly close to $1/2$. This behavior indicates that the effective central charge remains
close to the Ising-CFT value over the weak- and moderate-decoherence regime and the mixed state criticality between the mixed FM and PM is different from that of the compact boson CFT ($c=1$) well-verified in the qAT model\cite{Bridgeman2015,Chepiga2024}. 
Thus, we claim that for the mixed case, the existence of the phase boundary is the same as that of the qAT model, but the criticality is different in the two systems.

For $p_{zz}>0.3$, the extracted value of $c_{\rm eff}$ deviates from $1/2$ and suddenly drops. For $p_{zz}>0.4$, the value vanishes. 
This indicates that the surviving Ising CFT disappears and a different mixed state appears. 
Along with the behavior, we further plot the R2MI for $L_A=L/2$ in Fig.~\ref{Fig2}(c). 
The finite values of $MI^{(2)}(L_A)$ with a system size dependence for $0\leq p_{zz}\leq 0.3$ appears. For $p_{zz}\gtrapprox 0.3$, the value exhibits a sudden drop and settles down to a constant value $\ln 2$. That is, for a large $p_{zz}$ regime, the mixed state has a finite correlation.

The vanishing of the Ising CFT can correspond to the emergence of the SWSSB.
Therefore, the changes in the strong decoherence regime should be discussed through the R\'{e}nyi-2 $ZZ$ correlation as we do in the next section.

\begin{figure}[t]
\begin{center} 
\vspace{0.5cm}
\includegraphics[width=8cm]{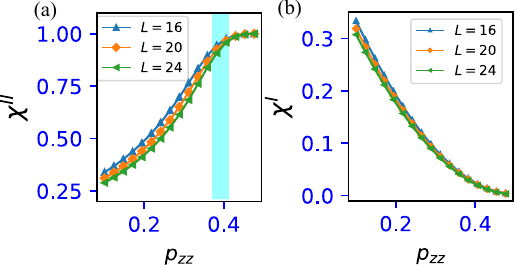}  
\end{center} 
\caption{$p_{zz}$-dependence of $\chi^{\rm II}_{ZZ}$[(a)] and $\chi^{\rm I}$[(b)]. 
From the right-blue regime in (a), the system size dependence vanishes and the values are saturated to almost unity.}
\label{Fig3}
\end{figure}

\subsection{R\'{e}nyi-2 $ZZ$ correlator}
To elucidate the behavior of the R2MI and its corresponding CFT, we here observe the susceptibilities for R\'{e}nyi-2 $ZZ$ correlator, $\chi^{\rm II}$ and conventional $ZZ$ correlator $\chi^{\rm I}$. 
These observables can characterize the emergence of the $Z_2$ SWSSB \cite{lessa2024} and are numerically tractable. 
These have already been observed in our previous study \cite{OKI2025_2}. In this work, we reproduce the results in the precise numerical conditions [especially, bond dimension $D=300$, etc.]. 

The reproduced results are shown in Figs.~\ref{Fig3} (a) and (b). 
In the regime $p_{zz}>0.4$, the SWSSB takes place, and as we observed in the above, the effective central charge $c_{\rm eff}$ vanishes. 
Within the accuracy and system sizes accessible in the present study, the loss of Ising criticality appears to rigorously coincide with the onset of SWSSB, although a narrow intermediate regime separating them cannot be conclusively excluded.

\begin{figure}[t]
\begin{center} 
\vspace{0.5cm}
\includegraphics[width=8.7cm]{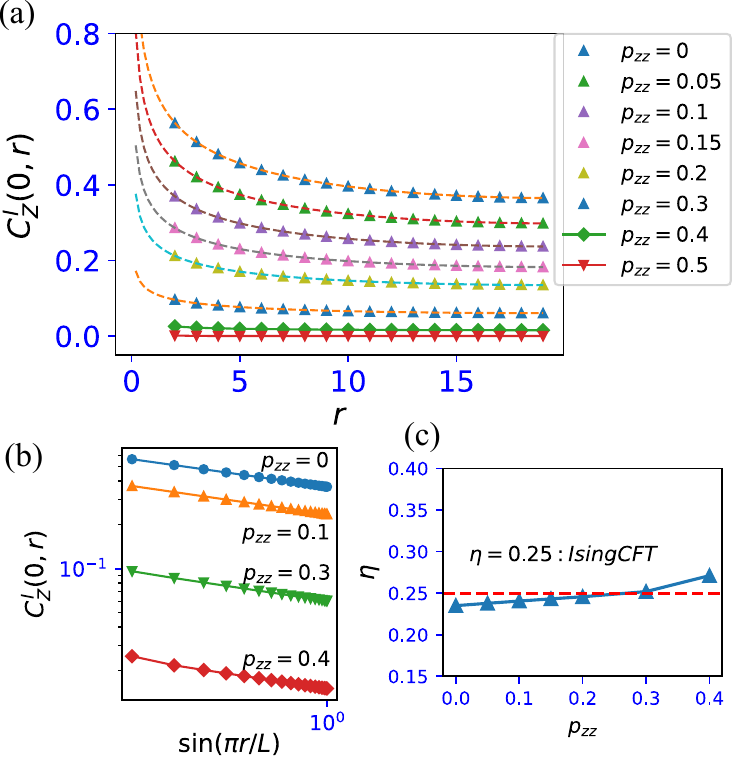}  
\end{center} 
\caption{Behavior of the correlation function $C^{I}_{Z}(0,r)$.
Data of $p_{zz}=0.05,0.1,0.15,0.2$ and $0.3$ are well-fitted by the power-law decay of Eq.(\ref{corr_zz_fit})(the extracted lines are represented by the dashed lines.), while the other data ($p_{zz}=0.4$ and $0.5$), the power-law decay is weakened. (b) 
Log-log plot of $C^{\mathrm{I}}_{Z}(0,r)$ as a function of
$\sin(\pi r/L)$ for representative values of $p_{zz}$.
The approximately linear behavior persists also at $p_{zz}=0.4$,
while the overall correlation amplitude is strongly suppressed. 
(c) Numerically extracted $\eta$ form the ansatz of Eq.(\ref{corr_zz_fit}).
We set $J=h=1$. The data are obtained for the $L=40$ system.
}
\label{Fig4}
\end{figure}
\subsection{Canonical $ZZ$ and $XX$ correlation functions}

In addition to the observation of the central charge in the preceding subsection, we shall show further evidence for the possible Ising critical properties of the mixed state 
under the decoherence. 

We observe the behavior of the canonical $ZZ$ correlation function defined by Eq.~(\ref{CIICI}). 
If the Ising CFT succeeds up to moderate-strength decoherence, the canonical $ZZ$ correlation function exhibits the property of the Ising CFT. 
In particular, the Ising CFT has the following form of the $ZZ$ correlation function, 
\begin{eqnarray}
C^{I}_{Z}(0,r)=\gamma_0\biggl(\frac{\pi}{L\sin (\pi r/L)}\biggr)^{\eta},
\label{corr_zz_fit0}
\end{eqnarray}
where the appearance of the sine function respects the chord distance $r$ due to the PBC. 
This sine-function part reflects the power law decay of the correlation, and $\eta$ is the conventional critical exponent of the spin correlation function.

Figure \ref{Fig4} (a) displays the $r$-dependence of $C^{I}_{Z}(0,r)$ for typical $p_{zz}$'s.
The mixed states on the phase boundary ($p_{zz}=0.05,0.1,0.15,0.2$ and $0.3$ data) clearly exhibit the power-law decay, while in the states for $p_{zz}=0.4$ and $0.5$ (the SWSSB regime), the correlation amplitude itself is strongly suppressed.  
The tendency of the power-law decay is supported by observing the log-log plot of the data of Fig.~\ref{Fig4}(a) as shown in Fig.~\ref{Fig4}(b). 
Here, we examine the detailed behavior; even for large $p_{zz}$, the behavior of power-law decay almost remains, 
but for large $p_{zz}$ the amplitude is very small. 
In particular, the linear tendency appears especially for $p_{zz}<0.4$. 
For large $p_{zz}$, the overall amplitude of the correlation function is strongly suppressed. 
As a result, a reliable quantitative determination of the critical exponent becomes difficult and sensitive. 
The identification of the large-$p_{zz}$ regime is therefore not based on this amplitude suppression alone, but on the other
observables discussed in Secs.~IV A and IV B.

Then, we extract the value of $\eta$ through the fitting ansatz in Eq.~(\ref{corr_zz_fit0}). The extraction of $\eta$ depends on the fitting range $i_0 \leq r \leq L/2$, particularly on the lower cutoff $i_0$,
because of short-distance nonuniversal corrections. For the results shown in Fig.~\ref{Fig4}(c), we employ the common
cutoff $i_0=2$ for all values of $p_{zz}$, independently of the resulting value of $\eta$.
Within the accessible system size, the extracted values remain close to the Ising-CFT value $\eta=1/4$ up to $p_{zz}\lessapprox 0.3$. We further examine the system-size dependence of the extracted
$\eta$ in Appendix~D using the same fixed cutoff $i_0^c=2$. Although the linearly extrapolated values exhibit small systematic deviations 
from the exact Ising-CFT value, the finite-size
dependence is weak and the estimates remain close to $\eta=1/4$. These results therefore indicate that the tendency toward Ising-like 
critical scaling is retained within the accessible system sizes.
However, we regard this analysis as a finite-size consistency check rather than as a high-precision determination of the asymptotic exponents.  
In passing, we also comment that $\gamma_0$ in
Eq.~(\ref{corr_zz_fit0}) decreases with increasing $p_{zz}$, which accounts for the decrease of $\chi^{\rm I}$ shown in Fig.~\ref{Fig3}(b). 

Here, we mention the behavior of the $ZZ$ spin correlation function on the critical line of the qAT model. 
The critical line is known as the orbifold CFT\cite{Kohmoto1981,O'Brien2015,Bridgeman2015}, where
the value of $\eta$ has been exactly obtained as $\eta=1/4$. 
As the value $\eta=1/4$ coincides with that of the Ising CFT, we cannot judge whether the above critical property of the decohered mixed state stems from the Ising or the qAT criticality.

\begin{figure}[t]
\begin{center} 
\vspace{0.5cm}
\includegraphics[width=8.7cm]{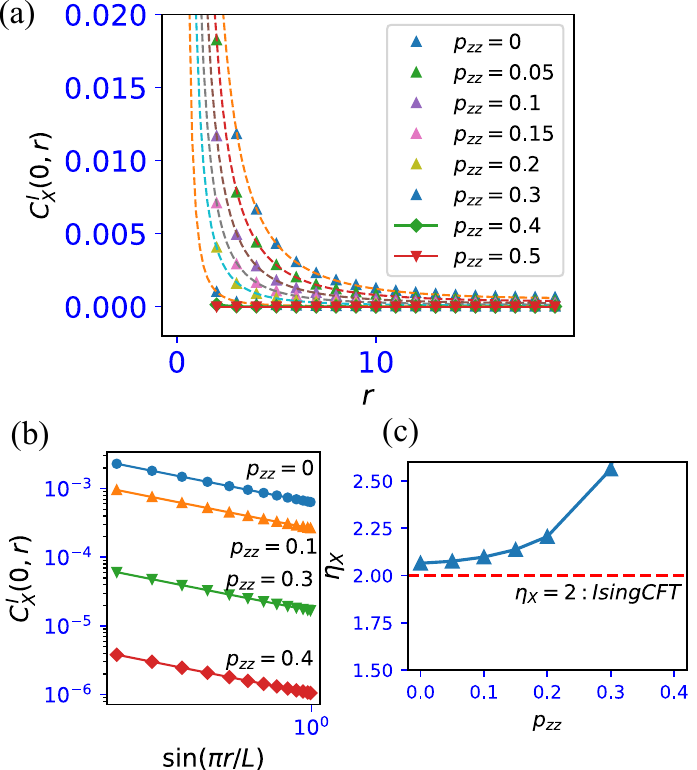}  
\end{center} 
\caption{Behavior of the correlation function $C^{I}_{X}(0,r)$[(a)]. Data of $p_{zz}=0.05,0.1,0.15$ and $0.2$ are well-fitted by 
the power-law decay (the dashed lines represent the fitting curves of Eq.(\ref{corr_zz_fit})), data of $p_{zz}=0.3$ exhibits a moderate deviation 
from $\eta_X=2$ even though the data is well-fitted by Eq.~(\ref{corr_zz_fit}) and the other data  $p_{zz}=0.4$ and $0.5$ deviates 
from $\eta_X=2$. 
(b) Log-log plot of $C^{I}_{X}(0,r)$ as a function of
$\sin(\pi r/L)$ for typical values of $p_{zz}$.
The approximately linear behavior persists even at $p_{zz}=0.4$,
while the overall correlation amplitude is strongly suppressed.
(c) Numerically extracted $\eta_X$ form the ansatz of Eq.(\ref{corr_zz_fit}).
We set $J=h=1$. System size $L=40$.}
\label{Fig8}
\end{figure}
Then, we further study its criticality by investigating the $XX$ correlation function, for which the Ising and qAT models exhibit different behavior. 
Here, we introduce the canonical connected $XX$ correlation function defined similarly to the $ZZ$ correlation function in Eq.~(\ref{CIICI}),
\begin{eqnarray}
C^{I}_{X}(i,j)&\equiv&\frac{\langle\langle {\bf 1}|X_{i,u}X_{j,u}|\rho_D\rangle\rangle}{\langle\langle {\bf 1}|\rho_D\rangle\rangle}\nonumber\\
&-&\frac{\langle\langle {\bf 1}|X_{i,u}|\rho_D\rangle\rangle}{\langle\langle {\bf 1}|\rho_D\rangle\rangle}\frac{\langle\langle {\bf 1}|X_{j,u}|\rho_D\rangle\rangle}{\langle\langle {\bf 1}|\rho_D\rangle\rangle}.
\end{eqnarray}
As the $ZZ$ correlation function, it is expected to obey the following form predicted by the CFT,
\begin{eqnarray}
C^{I}_{X}(0,r)=\gamma^X_0\biggl(\frac{\pi}{L\sin (\pi r/L)}\biggr)^{\eta_X}.
\label{corr_zz_fit}
\end{eqnarray}
This sine-function part reflects the power law decay of the correlation and $\eta_X$ is the critical exponent.
For the pure Ising critical state, it has been obtained that $\eta_X=2$, whereas for the qAT model, the value of $\eta_X$ is $\lambda$-dependent and given by $\eta_X=\pi/\arccos (-\lambda)$, 
a decreasing function of positive $\lambda$ and then $p_{zz}$.

The numerical results are shown in Fig.~\ref{Fig8} (a). 
Up to a moderate $p_{zz}$, $C^{I}_{X}(0,r)$ clearly exhibits a power-law decay with $\eta_X\sim 2.0$. The log-log plot of $C^{I}_{X}(0,r)$ supports this results as shown in Fig.~\ref{Fig8}(b). 
Again, we find that even for large $p_{zz}$ cases the power-law decay almost remains while in the states for $p_{zz}=0.4$ and $0.5$ (SWSSB regime), the overall correlation amplitude is strongly suppressed. 
This amplitude suppression alone does not distinguish the large-$p_{zz}$ regime, but it makes a reliable quantitative
determination of the critical exponent increasingly difficult and sensitive to the fitting range. We therefore next estimate $\eta_X$ in the regime where the
correlation amplitude remains sufficiently large. 

We next extract $\eta_X$ following the same procedure as for $\eta$. The numerically extracted values of $\eta_X$ are shown in Fig.~\ref{Fig8}(c), where we employ the common cutoff $i_0=2$ for 
all values of $p_{zz}$, independently of the resulting value of $\eta_X$.
Within the accessible system size, the extracted values are close to the Ising-CFT value $\eta_X=2$ for $p_{zz}\lessapprox 0.2$. 
We further examine the system-size dependence of the extracted $\eta_X$ in Appendix~D using the same fixed cutoff $i_0^c=2$. 
As in the case of $\eta$, the finite-size dependence is weak and the estimates remain close to the Ising-CFT value, although the linearly extrapolated values exhibit small systematic deviations from $\eta_X=2$.
We therefore regard this result as a finite-size consistency check of Ising-like critical scaling rather than as a high-precision determination of the asymptotic exponent.

We further verify whether the phase boundary line between the mixed FM and PM exhibits the properties of the Ising CFT. 
To obtain another evidence for that, we observe the behavior of the canonical $ZZ$ correlation function as varying $J$ in the TFIM with $h=1$, $p_{zz}$ fixed and $p_{x}=p_x=1/2-(1/2)(1-2p_{zz})^{1/J}$. 
This parameter sweep corresponds to orthogonally moving to the phase boundary line. 
Under this sweep, we calculate $C^{I}_{Z}(0,r)$ for various $J$ with $L$ fixed. 
Here, we expect that near the phase boundary, $C^{I}_{Z}(0,r)$ exhibits the following exponential decay,
\begin{eqnarray}
C^{I}_{Z}(0,r)= a_0\exp[-\frac{r}{\xi(J/h)}] + a_1,
\label{Czz_exp}
\end{eqnarray}
where $a_0$ and $a_1$ are non-universal constant and $\xi(J/h)$ is a correlation length depending on $J/h$. 
From the correlation length $\xi(J/h)$, we here define the correlation length critical exponent $\nu$ as $\xi(J/h)\propto |J/h-1|^{\nu}$. 

\begin{figure}[t]
\begin{center} 
\vspace{0.5cm}
\includegraphics[width=6.5cm]{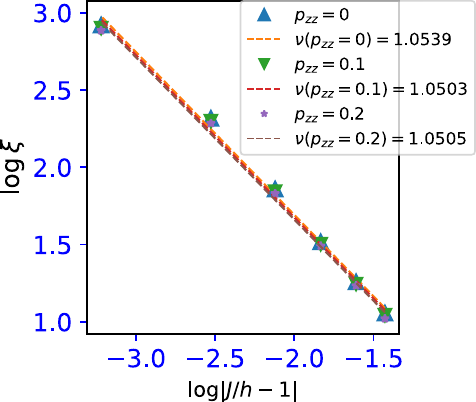}  
\end{center} 
\caption{Log-log plot for correlation length. The data of $C^{I}_{Z}(0,r)$ are calculated for $L=28$ system. The dashed lines are $\nu^* \log|J/h-1|+c_2$ where $\nu^*$ is an estimated value and $c_2$ is an offset.}
\label{Fig5}
\end{figure}

We estimate the correlation length $\xi(J/h)$ by fitting the data of $C^{I}_{Z}(0,r)$ to the form Eq.(\ref{Czz_exp}), where we use the range of $r$, [$3$,$L/2-1$] in $C^{I}_{Z}(0,r)$. 
That is, for the choice of the range of $r$, we exclude the short-distance region $r \le 2$, where non-universal small-scale effects exist. 
The upper bound $r \le L/2 - 1$ is chosen to avoid finite-size saturation effects due to periodic boundary conditions. 
Within this intermediate-distance window, the extracted exponent shows only weak dependence on moderate variations of the fitting range.   
Figure \ref{Fig5} shows the log-log plot of the extracted $\xi(J/h)$ vs $|J/h-1|$.  
We then carry out the fitting $\xi(J/h)\propto |J/h-1|^{\nu}$ for the data and estimate the value of $\nu$. 
We find that the extracted value $\nu$ for $p_{zz}\leq 0.2$ is much close to $1$ as shown in Fig.~\ref{Fig5}, 
independently to the value of $p_{zz}$.[In addition, we verified the estimated values of $\nu$ already converges in $L\geq 20$ (not shown).]
This means that the exponent is different from that of the qAT model (orbifold boson CFT), in which $\nu=1/[2-(\pi/2)(\arccos(-\lambda))^{-1}]$~\cite{Kohmoto1981}. 
In the Ising CFT, the exact correlation length critical exponent is $\nu=1$ \cite{di1996conformal}. 
Thus, the obtained result for the exponent $\nu$ is another evidence for the Ising CFT critical behavior of the mixed state under consideration. 

Taken all together, these numerical results indicate that the mixed phase-boundary line exhibits behavior 
consistent with Ising-like critical scaling within the accessible system sizes.

\section{Summary and conclusion} 
We investigated the fate of the Ising CFT on the critical Ising model under the $X+ZZ$ decoherence by employing the efficient numerical scheme based on the doubled Hilbert space formalism.
In the effective Hamiltonian picture, the decohered state is expected to possess a critical phase boundary up to a large decoherence regime. 
In the pure state picture by the qAT model, the boundary line is the orbifold boson CFT~\cite{Kohmoto1981,O'Brien2015}. 
On the other hand, the weak KW duality is respected in the target mixed system, and then, we expect that the pure Ising CFT can be preserved.

Our numerical results indicate that the critical behavior of the mixed-state phase boundary is consistent with Ising-like scaling within the accessible system sizes.
The subsystem dependence of the R2MI exhibits the expected CFT scaling form, and the extracted effective central charge remains close to the Ising-CFT value $c=1/2$ in the weak- and
moderate-decoherence regime. Moreover, the numerically extracted critical exponents remain close to the corresponding Ising-CFT values and exhibit only weak finite-size dependence.
We therefore regard these results as mutually consistent finite-size evidence for retained Ising-like criticality, rather than as a high-precision determination of the asymptotic CFT data. In the present doubled-state formulation, the system is described by a non-unitary mixed state generated through Kraus filterings. 
Consequently, we expect that the operator content required to stabilize the orbifold CFT is lost. Therefore, even on the critical line, the mixed state does not fall within the universality class of the orbifold CFT. 
On the other hand, the fact that the mixed-state phase structure mirrors that of the qAT model originates from the form of the filtering operation itself. In particular, the filtering can be expressed in an exponential form that resembles the qAT Hamiltonian. 
As a result, while the global phase diagram respects that of the qAT model, the critical behavior of the mixed states can be different from that of the genuine qAT model, as we showed in this study. We mention that in the previous studies \cite{Haegeman2015,Zhu2019} reporting a difference between the criticality, 
the filtering states exhibit different criticalities from those of the related effective Hamiltonian, 
obtained by approximately exponentiating the filtering operator. Concretely, this effective Hamiltonian is Toric code with 
magnetic perturbations \cite{Haegeman2015,Zhu2019}. 

We further investigate the threshold of the mixed criticality up to large decoherence strength.
From the behavior of the central charge numerically extracted by the CFT scaling ansatz, the central charge suddenly vanishes around $p_{zz}\sim 0.4$, 
indicating the mixed state Ising CFT is swept out and the critical mixed state is no longer stable. 
Then, the SWSSB characterized by R\'{e}nyi-2 $ZZ$ correlation supports the above behavior of the system. 
That is, vanishing the remaining Ising CFT corresponds to the appearance of the SWSSB. 
Our results indicate that Ising-like critical behavior persists along the mixed FM-PM phase boundary within the accessible system sizes. 
In contrast, the FM-SWSSB and PM-SWSSB transitions are decoherence-driven mixed-state transitions associated with SWSSB. 
Consequently, the universality classification based on CFT is not directly applicable in these regimes. A systematic finite-size scaling analysis of the corresponding critical exponents is left for future work.

\section*{Acknowledgements}
This work is supported by JSPS KAKENHI: JP23K13026 and JP26K06956(Y.K.) and JP23KJ0360 and JP26K17056(T.O.). 

\section*{Data availability}
The data that support the findings of this study are available from the authors upon reasonable request.\\

\renewcommand{\thesection}{A\arabic{section}} 
\renewcommand{\theequation}{A\arabic{equation}}
\renewcommand{\thefigure}{A\arabic{figure}}
\setcounter{equation}{0}
\setcounter{figure}{0}

\appendix
\section*{Appendix} 
\begin{figure}[b]
\begin{center} 
\vspace{0.5cm}
\includegraphics[width=6.5cm]{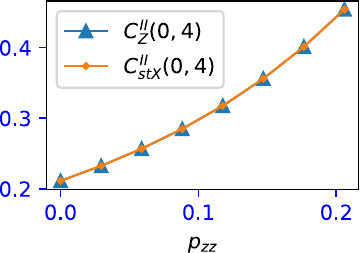}  
\end{center} 
\caption{Comparison between $C^{\rm II}_{ZZ}$ and $C^{\rm II}_{stX}$. 
We set $J=h=1$ and use $L=28$ system.}
\label{FigA1}
\end{figure}
\section{Calculation scheme for subsystem R\'{e}nyi-2 entanglement entropy}
We show a calculation scheme for the subsystem R\'{e}nyi-2 EE, where the subsystem is $X$.
The process of the partial trace for a subsystem $\bar{X}$ is carried out by the maximal depolarization channel \cite{Nielsen2011} represented by an operator $|\rho\rangle\rangle$ \cite{Lee2024}, which is given by
\begin{eqnarray}
\hat{D}_{\bar{X}}&=&\prod_{j\in \bar{X}}\hat{D}^{m}_j,\\
\hat{D}^{m}_j&=&\frac{1}{4}\biggr[\hat{I}_{j,u} \otimes \hat{I}_{j,\ell}+\hat{X}_{j,u} \otimes \hat{X}_{j,\ell}-\hat{Y}_{j,u}\otimes \hat{Y}_{j,\ell}\nonumber\\
&&+\hat{Z}_{j,u} \otimes \hat{Z}_{j,\ell}\biggr],
\end{eqnarray}
where $\hat{I}_{j,u(\ell)}$ is an identity operator for site-$j$ vector space in $\mathcal{H}_{u(\ell)}$, $Z(X,Y)_{j,u(\ell)}$ is Pauli-$Z$($X$,$Y$) operator at site $j$. 
The channel operator $\hat{D}_{\bar{X}}$ is a coupling between the system $u$ and $\ell$ in the ladder system. We apply $\hat{D}_{\bar{X}}$ to a state $|\rho\rangle\rangle$~\cite{zhang2025_SP}. Then, we obtain 
\begin{eqnarray}
\hat{D}_{\bar{X}}|\rho\rangle\rangle=|\frac{{\bf I}_{\bar{X}}}{d_{\bar{X}}}\otimes \rho_{X}\rangle\rangle,
\label{MD_1}
\end{eqnarray}
where $d_{\bar{X}}$ is the number of degree of freedom of the subsystem $\bar{X}$ and 
$|\frac{{\bf I}_{\bar{X}}}{d_{\bar{X}}}\otimes \rho_{X}\rangle\rangle \longleftrightarrow \frac{{\bf I}_{\bar{X}}}{d_{\bar{X}}}\otimes [{\rm Tr}_{\bar{X}}\rho]$. 

For the above vector, we consider the norm. Then, 
the norm of the state $\hat{D}_{\bar{X}}|\rho\rangle\rangle$ is related to $S^{(2)}_X$ as follows \cite{Ma2024_double},
\begin{eqnarray}
\langle\langle \frac{{\bf I}_{\bar{X}}}{d_{\bar{X}}}\otimes \rho_{X} |\frac{{\bf I}_{\bar{X}}}{d_{\bar{X}}}\otimes \rho_{X}\rangle\rangle 
=\frac{1}{d_{\bar{X}}}{\rm Tr}_{X}[\rho^2_{X}].
\label{norm_cal}
\end{eqnarray}
Thus, we obtain $S^{(2)}_X$ from  $\hat{D}_{\bar{X}}|\rho\rangle\rangle$ as
\begin{eqnarray}
S^{(2)}_X=-\log\biggr[d_{\bar{X}}\langle\langle \frac{{\bf I}_{\bar{X}}}{d_{\bar{X}}}\otimes \rho_{X} |\frac{{\bf I}_{\bar{X}}}{d_{\bar{X}}}\otimes \rho_{X}\rangle\rangle \biggl].
\end{eqnarray}

\begin{figure}[t]
\begin{center} 
\vspace{0.5cm}
\includegraphics[width=8cm]{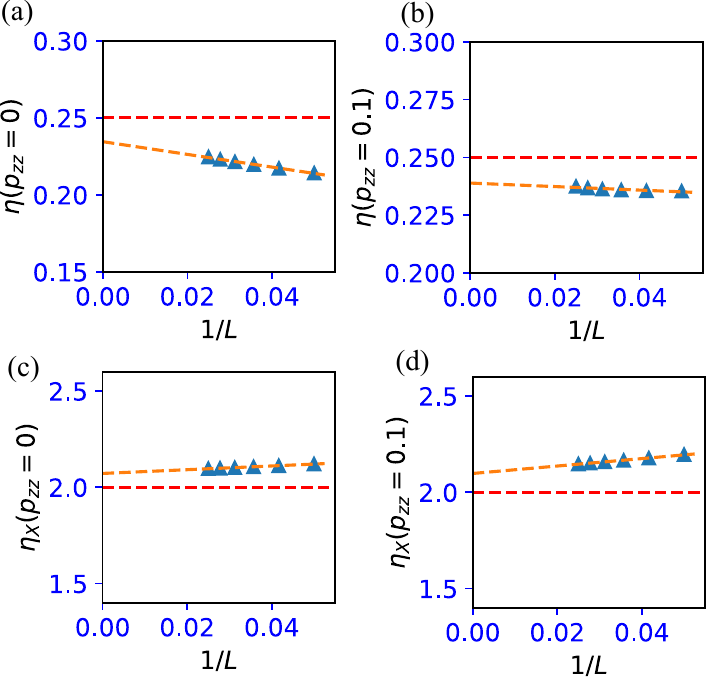}  
\end{center} 
\caption{
$1/L$ dependence of the numerically extracted exponents $\eta$ and $\eta_X$.
For all panels, we use the common lower cutoff $i_0^c=2$ and the fitting range $i_0^c\leq r\leq L/2$. (a) System-size dependence of $\eta$ for $p_{zz}=0$. (b) System-size dependence of $\eta$ for $p_{zz}=0.1$. (c) System-size dependence of $\eta_X$ for $p_{zz}=0$. (d) System-size dependence of $\eta_X$ for $p_{zz}=0.1$.
The orange dashed lines are linear fits in $1/L$, while the red dashed lines indicate the Ising-CFT values $\eta=1/4$ and
$\eta_X=2$. The extracted exponents exhibit only weak system-size dependence, although the extrapolated values show small systematic deviations from the exact Ising-CFT values.
}
\label{FigA2}
\end{figure}
The operation of Eq.~(\ref{MD_1}) and the norm calculation in Eq.~(\ref{norm_cal}) can be performed by the MPS numerical-based filtering methods.
Furthermore, we suitably set the subsystem $X$ and take a suitable combination. Then we obtain the R2MI, $MI^{(2)}(L_A)$.

\section{Canonical two point correlator from the doubled Hilbert space formalism}
For the 1D TFIM under the decoherence, the canonical two-point correlator is given by ${\rm Tr}[\rho_D Z_{i}Z_{j}]$. 
This observable is employed as an order parameter characterizing the ordinary long-range order. The expression in the doubled Hilbert space formalism is given by using the Choi isomorphism formula for a density matrix $\rho$ ~\cite{Choi1975,lee2023}. 
In general, the density matrix is represented by 
$$
|\rho\rangle\rangle=\frac{1}{\sqrt{{\rm dim}[\rho]}}\sum_{k}|k\rangle \otimes \rho|k\rangle,
$$
where $\{|k\rangle\}$ is a basis set on the single $\mathcal{H}$. 
Here, we consider an infinite temperature state, $\rho=\hat{I}/2^L$.
The state $\rho=\hat{I}/2^L$ can be regarded as a product state of the superposed triplet state with equal weight as shown in the main text, where we take the set of basis $\{|k\rangle\}$ as the spin $z$-component bases. 
Then, the canonical correlator ${\rm Tr}[\rho_D Z_{i}Z_{j}]$ for a decohered state $\rho_D$ is expressed as 
$$
{\rm Tr}[\rho_D Z_{i}Z_{j}]={\rm Tr}[\rho_D Z_{i}Z_{j}{\hat{I}}]=\frac{\langle\langle {\bf 1}|Z_{i,u}Z_{j,u}|\rho_D\rangle\rangle}{\langle\langle {\bf 1}|\rho_D\rangle\rangle}.
$$
The above quantity is $C^{I}_{Z}$ in the main text.

\section{Confirmation of Krammers-Wannier duality in weak symmetry sense}
We consider the KW transformed one of the $C^{\rm II}_{ZZ}(0,r)$ given by
$$
C^{\rm II}_{stX}(0,r)=\mbox{Tr}\biggl[(\prod_{0\leq\ell<r} X_\ell)\rho_D (\prod_{0\leq\ell<r} X_\ell) \rho_D\biggr].
$$
If the weak KW symmetry is sustained, then the equivalence, $C^{\rm II}_{ZZ}(0,r)=C^{\rm I}_{stX}(0,r)$ mentioned in the main text is expected. 

Figure \ref{FigA1} is the numerical result for the small $p_{zz}$ regime, where we fix $r=4$. 
We confirm that the equivalence is certainly satisfied. 
This means that the decoherence inherits the weakly symmetric KW dual, and the decohered critical state is so.

\section{Additional numerical estimation for the exponents $\eta$ and $\eta_X$}

In this Appendix, we show additional results for the critical exponents shown in Sec.IV. 
As mentioned in Sec.~IV, the numerical extraction of the exponents
$\eta$ and $\eta_X$ depends on the fitting range
$i_0 \leq r \leq L/2$, particularly on its lower bound, because the short-distance data are affected by nonuniversal corrections.
To avoid any ambiguity associated with the choice of the fitting window, we employ a common lower cutoff $i_0^c=2$ for all system
sizes, decoherence strengths, and correlation functions considered in this Appendix. 
Thus, the lower cutoff is fixed independently of $L$, $p_{zz}$, and the resulting value of the extracted exponent. 
This common fitting protocol also employed for Figs.~4(c) and 5(c) in the main text. 
Figure~\ref{FigA2} shows the system-size dependence of $\eta$ and $\eta_X$ extracted from the correlation functions $C^{I}_{Z}(0,r)$ and $C^{I}_{X}(0,r)$, respectively.
The orange dashed lines are linear fits in $1/L$ and are shown to illustrate the finite-size trends, whereas the red dashed lines indicate the exact Ising-CFT values. 
Figures~\ref{FigA2}(a) and \ref{FigA2}(b) show the $1/L$ dependence of the numerically extracted exponent $\eta$ for $p_{zz}=0$ and $0.1$, respectively. In both cases, the estimates exhibit only a weak and smooth system-size dependence over the accessible range.
The linear fits yield finite extrapolated values that are close to, but do not exhibit exact agreement with, the Ising-CFT value $\eta=1/4$.
The remaining deviation indicates systematic effects associated with the finite system sizes and the fixed fitting window.
We therefore do not interpret these linear fits as a high-precision determination of the thermodynamic-limit exponent.
Rather, the results show that the finite-size estimates remain stable and close to the Ising-CFT value under a common fixed-cutoff fitting protocol. 
Similarly, Figs.~\ref{FigA2}(c) and \ref{FigA2}(d) show the $1/L$ dependence of the numerically extracted exponent $\eta_X$ for $p_{zz}=0$ and $0.1$, respectively.
The estimates again display only a weak system-size dependence and show smooth trends toward finite extrapolated values.
These extrapolated values exhibit small systematic deviations from the exact Ising-CFT value $\eta_X=2$. 
Accordingly, we regard the results for both $\eta$ and $\eta_X$ as finite-size consistency checks of Ising-like scaling, rather than as precise determinations of the asymptotic critical exponents.

\bibliography{ref}

@article{Haegeman2015,
  title = {Gauging Quantum States: From Global to Local Symmetries in Many-Body Systems},
  author = {Haegeman, Jutho and Van Acoleyen, Karel and Schuch, Norbert and Cirac, J. Ignacio and Verstraete, Frank},
  journal = {Phys. Rev. X},
  volume = {5},
  issue = {1},
  pages = {011024},
  numpages = {10},
  year = {2015},
  month = {Feb},
  publisher = {American Physical Society},
  doi = {10.1103/PhysRevX.5.011024},
  url = {https://link.aps.org/doi/10.1103/PhysRevX.5.011024}
}

@article{Kohmoto1981,
  title = {Hamiltonian studies of the $d=2$ Ashkin-Teller model},
  author = {Kohmoto, Mahito and den Nijs, Marcel and Kadanoff, Leo P.},
  journal = {Phys. Rev. B},
  volume = {24},
  issue = {9},
  pages = {5229--5241},
  numpages = {0},
  year = {1981},
  month = {Nov},
  publisher = {American Physical Society},
  doi = {10.1103/PhysRevB.24.5229},
  url = {https://link.aps.org/doi/10.1103/PhysRevB.24.5229}
}

@article{lee2023,
  title = {Quantum Criticality Under Decoherence or Weak Measurement},
  author = {Lee, Jong Yeon and Jian, Chao-Ming and Xu, Cenke},
  journal = {Phys. Rev. X Quantum},
  volume = {4},
  issue = {3},
  pages = {030317},
  numpages = {20},
  year = {2023},
  month = {Aug},
  publisher = {American Physical Society},
  doi = {10.1103/PRXQuantum.4.030317},
  url = {https://link.aps.org/doi/10.1103/PRXQuantum.4.030317}
}

@article{Zhu2019,
  title = {Gapless Coulomb State Emerging from a Self-Dual Topological Tensor-Network State},
  author = {Zhu, Guo-Yi and Zhang, Guang-Ming},
  journal = {Phys. Rev. Lett.},
  volume = {122},
  issue = {17},
  pages = {176401},
  numpages = {6},
  year = {2019},
  month = {Apr},
  publisher = {American Physical Society},
  doi = {10.1103/PhysRevLett.122.176401},
  url = {https://link.aps.org/doi/10.1103/PhysRevLett.122.176401}
}

@article{Chen2024_v2,
  title = {Unconventional topological mixed-state transition and critical phase induced by self-dual coherent errors},
  author = {Chen, Yu-Hsueh and Grover, Tarun},
  journal = {Phys. Rev. B},
  volume = {110},
  issue = {12},
  pages = {125152},
  numpages = {17},
  year = {2024},
  month = {Sep},
  publisher = {American Physical Society},
  doi = {10.1103/PhysRevB.110.125152},
  url = {https://link.aps.org/doi/10.1103/PhysRevB.110.125152}
}

@article{Lee2024,
  doi = {10.22331/q-2025-01-23-1607},
  url = {https://doi.org/10.22331/q-2025-01-23-1607},
  title = {Symmetry protected topological phases under decoherence},
  author = {Lee, Jong Yeon and You, Yi-Zhuang and Xu, Cenke},
  journal = {{Quantum}},
  issn = {2521-327X},
  publisher = {{Verein zur F{\"{o}}rderung des Open Access Publizierens in den Quantenwissenschaften}},
  volume = {9},
  pages = {1607},
  month = jan,
  year = {2025}
}

@article{Castelnovo2008,
  title = {Quantum topological phase transition at the microscopic level},
  author = {Castelnovo, Claudio and Chamon, Claudio},
  journal = {Phys. Rev. B},
  volume = {77},
  issue = {5},
  pages = {054433},
  numpages = {14},
  year = {2008},
  month = {Feb},
  publisher = {American Physical Society},
  doi = {10.1103/PhysRevB.77.054433},
  url = {https://link.aps.org/doi/10.1103/PhysRevB.77.054433}
}

@article{Choi1975,
title = {Completely positive linear maps on complex matrices},
journal = {Lin. Alg. Appl.},
volume = {10},
number = {3},
pages = {285-290},
year = {1975},
issn = {0024-3795},
doi = {https://doi.org/10.1016/0024-3795(75)90075-0},
url = {https://www.sciencedirect.com/science/article/pii/0024379575900750},
author = {Man-Duen Choi}
}

@article{JAMIOLKOWSKI1972,
title = {Linear transformations which preserve trace and positive semidefiniteness of operators},
journal = {Rep. Math. Phys.},
volume = {3},
number = {4},
pages = {275-278},
year = {1972},
issn = {0034-4877},
doi = {https://doi.org/10.1016/0034-4877(72)90011-0},
url = {https://www.sciencedirect.com/science/article/pii/0034487772900110},
author = {A. Jamiołkowski}
}

@book{Nielsen2011,
author = {Nielsen, Michael A. and Chuang, Isaac L.},
title = {Quantum Computation and Quantum Information},
year = {2011},
isbn = {1107002176},
publisher = {Cambridge University Press},
address = {USA},
edition = {10th}
}

@Article{TeNPy,
	title={{Efficient numerical simulations with Tensor Networks: Tensor Network Python (TeNPy)}},
	author={Johannes Hauschild and Frank Pollmann},
	journal={SciPost Phys. Lect. Notes},
	pages={5},
	year={2018},
	publisher={SciPost},
	doi={10.21468/SciPostPhysLectNotes.5},
	url={https://scipost.org/10.21468/SciPostPhysLectNotes.5},
}

@Article{Hauschild2024,
	title={{Tensor network Python (TeNPy) version 1}},
	author={Johannes Hauschild and Jakob Unfried and Sajant Anand and Bartholomew Andrews and Marcus Bintz and Umberto Borla and Stefan Divic and Markus Drescher and Jan Geiger and Martin Hefel and Kévin Hémery and Wilhelm Kadow and Jack Kemp and Nico Kirchner and Vincent S. Liu and Gunnar Möller and Daniel Parker and Michael Rader and Anton Romen and Samuel Scalet and Leon Schoonderwoerd and Maximilian Schulz and Tomohiro Soejima and Philipp Thoma and Yantao Wu and Philip Zechmann and Ludwig Zweng and Roger S. K. Mong and Michael P. Zaletel and Frank Pollmann},
	journal={SciPost Phys. Codebases},
	pages={41},
	year={2024},
	publisher={SciPost},
	doi={10.21468/SciPostPhysCodeb.41},
	url={https://scipost.org/10.21468/SciPostPhysCodeb.41},
}

@article{groot2022,
   title={Symmetry Protected Topological Order in Open Quantum Systems},
   volume={6},
   ISSN={2521-327X},
   url={http://dx.doi.org/10.22331/q-2022-11-10-856},
   DOI={10.22331/q-2022-11-10-856},
   journal={Quantum},
   publisher={Verein zur Forderung des Open Access Publizierens in den Quantenwissenschaften},
   author={de Groot, Caroline and Turzillo, Alex and Schuch, Norbert},
   year={2022},
   month=nov, pages={856} 
}

@article{sala2024,
  title = {Spontaneous strong symmetry breaking in open systems: Purification perspective},
  author = {Sala, Pablo and Gopalakrishnan, Sarang and Oshikawa, Masaki and You, Yizhi},
  journal = {Phys. Rev. B},
  volume = {110},
  issue = {15},
  pages = {155150},
  numpages = {28},
  year = {2024},
  month = {Oct},
  publisher = {American Physical Society},
  doi = {10.1103/PhysRevB.110.155150},
  url = {https://link.aps.org/doi/10.1103/PhysRevB.110.155150}
}

@article{lessa2024,
  title = {Strong-to-Weak Spontaneous Symmetry Breaking in Mixed Quantum States},
  author = {Lessa, Leonardo A. and Ma, Ruochen and Zhang, Jian-Hao and Bi, Zhen and Cheng, Meng and Wang, Chong},
  journal = {PRX Quantum},
  volume = {6},
  issue = {1},
  pages = {010344},
  numpages = {24},
  year = {2025},
  month = {Mar},
  publisher = {American Physical Society},
  doi = {10.1103/PRXQuantum.6.010344},
  url = {https://link.aps.org/doi/10.1103/PRXQuantum.6.010344}
}

@article{Ashkin1943,
  title = {Statistics of Two-Dimensional Lattices with Four Components},
  author = {Ashkin, J. and Teller, E.},
  journal = {Phys. Rev.},
  volume = {64},
  issue = {5-6},
  pages = {178--184},
  numpages = {0},
  year = {1943},
  month = {Sep},
  publisher = {American Physical Society},
  doi = {10.1103/PhysRev.64.178},
  url = {https://link.aps.org/doi/10.1103/PhysRev.64.178}
}

@article{Solyom,
  title = {Duality of the block transformation and decimation for quantum spin systems},
  author = {S\'olyom, J.},
  journal = {Phys. Rev. B},
  volume = {24},
  issue = {1},
  pages = {230--243},
  numpages = {0},
  year = {1981},
  month = {Jul},
  publisher = {American Physical Society},
  doi = {10.1103/PhysRevB.24.230},
  url = {https://link.aps.org/doi/10.1103/PhysRevB.24.230}
}

@book{gardiner2000,
  author = {Gardiner, C. W. and Zoller, P.},
  edition = {second},
  editor = {Haken, H.},
  publisher = {Springer},
  title = {Quantum Noise},
  year = 2000
}

@article{preskill2018,
   title={Quantum Computing in the NISQ era and beyond},
   volume={2},
   ISSN={2521-327X},
   url={http://dx.doi.org/10.22331/q-2018-08-06-79},
   DOI={10.22331/q-2018-08-06-79},
   journal={Quantum},
   publisher={Verein zur Forderung des Open Access Publizierens in den Quantenwissenschaften},
   author={Preskill, John},
   year={2018},
   month=aug, pages={79} }

@article{dennis2002,
   title={Topological quantum memory},
   volume={43},
   ISSN={1089-7658},
   url={http://dx.doi.org/10.1063/1.1499754},
   DOI={10.1063/1.1499754},
   number={9},
   journal={J. Math. Phys.},
   publisher={AIP Publishing},
   author={Dennis, Eric and Kitaev, Alexei and Landahl, Andrew and Preskill, John},
   year={2002},
   month=sep, pages={4452–4505} 
}

@article{wang2003,
   title={Confinement-Higgs transition in a disordered gauge theory and the accuracy threshold for quantum memory},
   volume={303},
   ISSN={0003-4916},
   url={http://dx.doi.org/10.1016/S0003-4916(02)00019-2},
   DOI={10.1016/s0003-4916(02)00019-2},
   number={1},
   journal={Ann.Phys.},
   publisher={Elsevier BV},
   author={Wang, Chenyang and Harrington, Jim and Preskill, John},
   year={2003},
   month=jan, pages={31–58} }

@Article{bluvstein2024,
author={Bluvstein, Dolev
and Evered, Simon J.
and Geim, Alexandra A.
and Li, Sophie H.
and Zhou, Hengyun
and Manovitz, Tom
and Ebadi, Sepehr
and Cain, Madelyn
and Kalinowski, Marcin
and Hangleiter, Dominik
and Bonilla Ataides, J. Pablo
and Maskara, Nishad
and Cong, Iris
and Gao, Xun
and Sales Rodriguez, Pedro
and Karolyshyn, Thomas
and Semeghini, Giulia
and Gullans, Michael J.
and Greiner, Markus
and Vladan, Vuletiﾄ・
and Lukin, Mikhail D.},
title={Logical quantum processor based on reconfigurable atom arrays},
journal={Nature},
year={2024},
month={Feb},
day={01},
volume={626},
number={7997},
pages={58-65},
issn={1476-4687},
doi={10.1038/s41586-023-06927-3},
url={https://doi.org/10.1038/s41586-023-06927-3}
}

@Article{ebadi2021,
author={Ebadi, Sepehr
and Wang, Tout T.
and Levine, Harry
and Keesling, Alexander
and Semeghini, Giulia
and Omran, Ahmed
and Bluvstein, Dolev
and Samajdar, Rhine
and Pichler, Hannes
and Ho, Wen Wei
and Choi, Soonwon
and Sachdev, Subir
and Greiner, Markus
and Vladan, Vuletiﾄ・
and Lukin, Mikhail D.},
title={Quantum phases of matter on a 256-atom programmable quantum simulator},
journal={Nature},
year={2021},
month={Jul},
day={01},
volume={595},
number={7866},
pages={227-232},
issn={1476-4687},
doi={10.1038/s41586-021-03582-4},
url={https://doi.org/10.1038/s41586-021-03582-4}
}

@article{Ardonne2004,
title = {Topological order and conformal quantum critical points},
journal = {Ann.Phys.},
volume = {310},
number = {2},
pages = {493-551},
year = {2004},
issn = {0003-4916},
doi = {https://doi.org/10.1016/j.aop.2004.01.004},
url = {https://www.sciencedirect.com/science/article/pii/S0003491604000247},
author = {Eddy Ardonne and Paul Fendley and Eduardo Fradkin}
}

@article{ohno2004,
title = {Phase structure of the random-plaquette Z2 gauge model: accuracy threshold for a toric quantum memory},
journal = {Nucl. Phys. B},
volume = {697},
number = {3},
pages = {462-480},
year = {2004},
issn = {0550-3213},
doi = {https://doi.org/10.1016/j.nuclphysb.2004.07.003},
url = {https://www.sciencedirect.com/science/article/pii/S055032130400481X},
author = {Takuya Ohno and Gaku Arakawa and Ikuo Ichinose and Tetsuo Matsui}
}

@article{Bridgeman2015,
  title = {Multiscale entanglement renormalization ansatz for spin chains with continuously varying criticality},
  author = {Bridgeman, Jacob C. and O'Brien, Aroon and Bartlett, Stephen D. and Doherty, Andrew C.},
  journal = {Phys. Rev. B},
  volume = {91},
  issue = {16},
  pages = {165129},
  numpages = {13},
  year = {2015},
  month = {Apr},
  publisher = {American Physical Society},
  doi = {10.1103/PhysRevB.91.165129},
  url = {https://link.aps.org/doi/10.1103/PhysRevB.91.165129}
}

@article{wang2024,
  title = {Intrinsic Mixed-State Topological Order},
  author = {Wang, Zijian and Wu, Zhengzhi and Wang, Zhong},
  journal = {PRX Quantum},
  volume = {6},
  issue = {1},
  pages = {010314},
  numpages = {29},
  year = {2025},
  month = {Jan},
  publisher = {American Physical Society},
  doi = {10.1103/PRXQuantum.6.010314},
  url = {https://link.aps.org/doi/10.1103/PRXQuantum.6.010314}
}

@article{sohal2024,
  title = {Noisy Approach to Intrinsically Mixed-State Topological Order},
  author = {Sohal, Ramanjit and Prem, Abhinav},
  journal = {PRX Quantum},
  volume = {6},
  issue = {1},
  pages = {010313},
  numpages = {20},
  year = {2025},
  month = {Jan},
  publisher = {American Physical Society},
  doi = {10.1103/PRXQuantum.6.010313},
  url = {https://link.aps.org/doi/10.1103/PRXQuantum.6.010313}
}

@article{CASTELNOVO2005,
   title={From quantum mechanics to classical statistical physics: Generalized Rokhsar–Kivelson Hamiltonians and the “Stochastic Matrix Form” decomposition},
   volume={318},
   ISSN={0003-4916},
   url={http://dx.doi.org/10.1016/j.aop.2005.01.006},
   DOI={10.1016/j.aop.2005.01.006},
   number={2},
   journal={Ann.Phys.},
   publisher={Elsevier BV},
   author={Castelnovo, Claudio and Chamon, Claudio and Mudry, Christopher and Pujol, Pierre},
   year={2005},
   month=aug, pages={316–344} }

@article{Yamanaka1994,
  title = {Phase diagram of the Ashkin-Teller quantum spin chain},
  author = {Yamanaka, Masanori and Hatsugai, Yasuhiro and Kohmoto, Mahito},
  journal = {Phys. Rev. B},
  volume = {50},
  issue = {1},
  pages = {559--562},
  numpages = {0},
  year = {1994},
  month = {Jul},
  publisher = {American Physical Society},
  doi = {10.1103/PhysRevB.50.559},
  url = {https://link.aps.org/doi/10.1103/PhysRevB.50.559}
}

@article{KOI2024_IMTO,
  title = {Intrinsic mixed-state topological order in a stabilizer system under stochastic decoherence: Strong-to-weak spontaneous symmetry breaking from a percolation point of view},
  author = {Kuno, Yoshihito and Orito, Takahiro and Ichinose, Ikuo},
  journal = {Phys. Rev. B},
  volume = {111},
  issue = {6},
  pages = {064111},
  numpages = {12},
  year = {2025},
  month = {Feb},
  publisher = {American Physical Society},
  doi = {10.1103/PhysRevB.111.064111},
  url = {https://link.aps.org/doi/10.1103/PhysRevB.111.064111}
}

@article{CC2004,
doi = {10.1088/1742-5468/2004/06/P06002},
url = {https://dx.doi.org/10.1088/1742-5468/2004/06/P06002},
year = {2004},
month = {jun},
publisher = {},
volume = {2004},
number = {06},
pages = {P06002},
author = {Pasquale Calabrese and  John Cardy},
title = {Entanglement entropy and quantum field theory},
journal = {J. Stat. Mech.}
}

@article{Guo-and-Ashida2024,
  title = {Two-dimensional symmetry-protected topological phases and transitions in open quantum systems},
  author = {Guo, Yuxuan and Ashida, Yuto},
  journal = {Phys. Rev. B},
  volume = {109},
  issue = {19},
  pages = {195420},
  numpages = {10},
  year = {2024},
  month = {May},
  publisher = {American Physical Society},
  doi = {10.1103/PhysRevB.109.195420},
  url = {https://link.aps.org/doi/10.1103/PhysRevB.109.195420}
}

@misc{Zurek2003decoherence,
      title={Decoherence and the transition from quantum to classical -- REVISITED}, 
      author={Wojciech H. Zurek},
      year={2003},
      eprint={quant-ph/0306072},
      archivePrefix={arXiv},
      primaryClass={quant-ph},
      url={https://arxiv.org/abs/quant-ph/0306072}, 
}

@book{di1996conformal,
  title={Conformal Field Theory},
  author={Di Francesco, P. and Mathieu, P. and S{\'e}n{\'e}chal, D.},
  isbn={9781461222576},
  series={Graduate texts in contemporary physics},
  url={https://books.google.co.jp/books?id=mcMbswEACAAJ},
  year={1996},
  publisher={Island Press}
}

@book{Henkel1999conformal,
  title={Conformal Invariance and Critical Phenomena},
  author={Henkel, M.},
  isbn={9783540653219},
  lccn={99011044},
  series={Texts and monographs in physics},
  url={https://books.google.co.jp/books?id=oobtgkfSBdcC},
  year={1999},
  publisher={Springer}
}

@article{Zou2023,
  title = {Channeling Quantum Criticality},
  author = {Zou, Yijian and Sang, Shengqi and Hsieh, Timothy H.},
  journal = {Phys. Rev. Lett.},
  volume = {130},
  issue = {25},
  pages = {250403},
  numpages = {7},
  year = {2023},
  month = {Jun},
  publisher = {American Physical Society},
  doi = {10.1103/PhysRevLett.130.250403},
  url = {https://link.aps.org/doi/10.1103/PhysRevLett.130.250403}
}

@article{Fan_2024,
   title={Diagnostics of Mixed-State Topological Order and Breakdown of Quantum Memory},
   volume={5},
   ISSN={2691-3399},
   url={http://dx.doi.org/10.1103/PRXQuantum.5.020343},
   DOI={10.1103/prxquantum.5.020343},
   number={2},
   journal={PRX Quantum},
   publisher={American Physical Society (APS)},
   author={Fan, Ruihua and Bao, Yimu and Altman, Ehud and Vishwanath, Ashvin},
   year={2024},
   month=may
}

@article{KOI2025_6,
  title = {R\'enyi Markov length in one-dimensional nontrivial mixed-state phases},
  author = {Kuno, Yoshihito and Orito, Takahiro and Ichinose, Ikuo},
  journal = {Phys. Rev. B},
  volume = {113},
  issue = {1},
  pages = {014113},
  numpages = {12},
  year = {2026},
  month = {Jan},
  publisher = {American Physical Society},
  doi = {10.1103/y8y1-2gqc},
  url = {https://link.aps.org/doi/10.1103/y8y1-2gqc}
}

@article{OKI2025_2,
  title = {Strong and weak symmetries and their spontaneous symmetry breaking in mixed states emerging from the quantum Ising model under multiple decoherence},
  author = {Orito, Takahiro and Kuno, Yoshihito and Ichinose, Ikuo},
  journal = {Phys. Rev. B},
  volume = {111},
  issue = {5},
  pages = {054106},
  numpages = {11},
  year = {2025},
  month = {Feb},
  publisher = {American Physical Society},
  doi = {10.1103/PhysRevB.111.054106},
  url = {https://link.aps.org/doi/10.1103/PhysRevB.111.054106}
}

@article{Stephan2010,
  title = {R\'enyi entropy of a line in two-dimensional Ising models},
  author = {St\'ephan, J.-M. and Misguich, G. and Pasquier, V.},
  journal = {Phys. Rev. B},
  volume = {82},
  issue = {12},
  pages = {125455},
  numpages = {8},
  year = {2010},
  month = {Sep},
  publisher = {American Physical Society},
  doi = {10.1103/PhysRevB.82.125455},
  url = {https://link.aps.org/doi/10.1103/PhysRevB.82.125455}
}

@article{Alcaraz2013,
  title = {Universal Behavior of the Shannon Mutual Information of Critical Quantum Chains},
  author = {Alcaraz, F. C. and Rajabpour, M. A.},
  journal = {Phys. Rev. Lett.},
  volume = {111},
  issue = {1},
  pages = {017201},
  numpages = {5},
  year = {2013},
  month = {Jul},
  publisher = {American Physical Society},
  doi = {10.1103/PhysRevLett.111.017201},
  url = {https://link.aps.org/doi/10.1103/PhysRevLett.111.017201}
}

@article{Alcaraz2014,
  title = {Universal behavior of the Shannon and R\'enyi mutual information of quantum critical chains},
  author = {Alcaraz, F. C. and Rajabpour, M. A.},
  journal = {Phys. Rev. B},
  volume = {90},
  issue = {7},
  pages = {075132},
  numpages = {10},
  year = {2014},
  month = {Aug},
  publisher = {American Physical Society},
  doi = {10.1103/PhysRevB.90.075132},
  url = {https://link.aps.org/doi/10.1103/PhysRevB.90.075132}
}

@article{Stephan2014,
  title = {Shannon and R\'enyi mutual information in quantum critical spin chains},
  author = {St\'ephan, Jean-Marie},
  journal = {Phys. Rev. B},
  volume = {90},
  issue = {4},
  pages = {045424},
  numpages = {18},
  year = {2014},
  month = {Jul},
  publisher = {American Physical Society},
  doi = {10.1103/PhysRevB.90.045424},
  url = {https://link.aps.org/doi/10.1103/PhysRevB.90.045424}
}

@article{LinhaoLi2025,
	title={{Intrinsically/purely gapless-SPT from non-invertible duality transformations}},
	author={Linhao Li and Masaki Oshikawa and Yunqin Zheng},
	journal={SciPost Phys.},
	volume={18},
	pages={153},
	year={2025},
	publisher={SciPost},
	doi={10.21468/SciPostPhys.18.5.153},
	url={https://scipost.org/10.21468/SciPostPhys.18.5.153},
}

@misc{Note_deco,
note = 
{Note that the order in application of the above local channels is irrelevant as long as we consider decoherence channels using Pauli operators such as $\mathcal{E}_{g_j}[\rho]=(1-p)\rho+g_j \rho g^\dagger_j$, where $g_j$ is an element of Pauli group with a finite length support.
Then, general two channels are commutative, $\mathcal{E}_{g_j}\circ \mathcal{E}_{g_\ell}[\rho]=\mathcal{E}_{g_\ell}\circ \mathcal{E}_{g_j}[\rho]$ for either $[g_j,g_\ell]=0$ or $\{g_j,g_\ell\}=0$.
}
}

@misc{Coef_nomal,
note = 
{The coefficient of the last line in Eq.(3) is given by $C(p_{zz},p_x,L)\equiv (1-2p_{zz})^{L/2}(1-2p_x)^{L/2}$. The norm $\langle\langle \rho_D|\rho_D\rangle\rangle$ corresponds to the purity $\Tr[\rho_D^2]$ ($>0$), and the norm exhibits an exponential decay with the system size $L$ due to the factor $C(p_{zz},p_x,L)$.
}
}

@misc{Filtering,
note = 
{This effective Hamiltonian picture has succeeded in obtaining states close to the true ground states in various perturbed Hamiltonians such as Toric code \cite{Ardonne2004,CASTELNOVO2005,Castelnovo2008,Haegeman2015}.
}
}

@misc{KOI_2025_v4,
      title={R\'{e}nyi Markov length in one-dimensional non-trivial mixed state phases and mixed state phase transitions}, 
      author={Yoshihito Kuno and Takahiro Orito and Ikuo Ichinose},
      year={2025},
      eprint={2505.02125},
      archivePrefix={arXiv},
      primaryClass={quant-ph},
      url={https://arxiv.org/abs/2505.02125}, 
}

@article{zhang2025_SP,
  title = {Universal Properties of Critical Mixed-States from Measurement and Feedback},
  author = {Zhang, Zhehao and Zou, Yijian and Hsieh, Timothy H. and Vijay, Sagar},
  journal = {PRX Quantum},
  volume = {6},
  issue = {4},
  pages = {040363},
  numpages = {29},
  year = {2025},
  month = {Dec},
  publisher = {American Physical Society},
  doi = {10.1103/66zv-8k1t},
  url = {https://link.aps.org/doi/10.1103/66zv-8k1t}
}

@misc{Ma2024_double,
      title={Symmetry Protected Topological Phases of Mixed States in the Doubled Space}, 
      author={Ruochen Ma and Alex Turzillo},
      year={2024},
      eprint={2403.13280},
      archivePrefix={arXiv},
      primaryClass={quant-ph},
      url={https://arxiv.org/abs/2403.13280}, 
}

@article{AFFLECK199735,
title = {Boundary condition changing operations in conformal field theory and condensed matter physics},
journal = {Nuclear Physics B - Proceedings Supplements},
volume = {58},
pages = {35-41},
year = {1997},
note = {Proceedings of the European Research Conference in the Memory of Claude Itzykson},
issn = {0920-5632},
doi = {https://doi.org/10.1016/S0920-5632(97)00411-8},
url = {https://www.sciencedirect.com/science/article/pii/S0920563297004118},
author = {Ian Affleck}
}

@article{KOI2025_ZZ_deco,
	title = {System-environmental entanglement in critical spin systems under $ZZ$-decoherence and its relation to strong and weak symmetries},
	pages = {014},
	author = {Kuno, Yoshihito and Orito, Takahiro and Ichinose, Ikuo},
	journal = {SciPost Phys. Core},
	volume = {9},
	year = {2026},
	publisher = {SciPost},
	doi = {10.21468/SciPostPhysCore.9.1.014},
	url = {https://scipost.org/10.21468/SciPostPhysCore.9.1.014}
}

@article{Chepiga2024,
  title = {Probing universal critical scaling with scan density matrix renormalization group},
  author = {Chepiga, Natalia},
  journal = {Phys. Rev. B},
  volume = {110},
  issue = {14},
  pages = {144401},
  numpages = {11},
  year = {2024},
  month = {Oct},
  publisher = {American Physical Society},
  doi = {10.1103/PhysRevB.110.144401},
  url = {https://link.aps.org/doi/10.1103/PhysRevB.110.144401}
}

@article{PhysRevB.99.224432,
  title = {Dissipative quantum Ising chain as a non-Hermitian Ashkin-Teller model},
  author = {Shibata, Naoyuki and Katsura, Hosho},
  journal = {Phys. Rev. B},
  volume = {99},
  issue = {22},
  pages = {224432},
  numpages = {8},
  year = {2019},
  month = {Jun},
  publisher = {American Physical Society},
  doi = {10.1103/PhysRevB.99.224432},
  url = {https://link.aps.org/doi/10.1103/PhysRevB.99.224432}
}

@misc{footnote1,
note = 
{We note that for the full system, Eq.~(\ref{S2_con}) is not applied with $L_A=L$, but we assume $S^{(2)}_A(L_A)=\alpha L + \beta_{L}$. 
Then the nonuniversal contribution $\beta_L$ is absorbed into the constant $\beta_1$.}
}

\end{document}